\documentclass{article}

\usepackage{arxiv}

\usepackage[utf8]{inputenc} % allow utf-8 input
\usepackage[T1]{fontenc}    % use 8-bit T1 fonts
\usepackage{hyperref}       % hyperlinks
\usepackage{url}            % simple URL typesetting
\usepackage{booktabs}       % professional-quality tables
\usepackage{amsfonts}       % blackboard math symbols
\usepackage{nicefrac}       % compact symbols for 1/2, etc.
\usepackage{microtype}      % microtypography
\usepackage{lipsum}
\usepackage{graphicx}
\usepackage{amsmath}
\usepackage{appendix}
\graphicspath{ {./images/} }

\title{Learning Collective Dynamics of Multi-Agent Systems \\ using Event-based Vision}

\author{
 Minah Lee, Uday Kamal, and Saibal Mukhopadhyay \\
  School of Electrical and Computer Engineering\\
  Georgia Institute of Technology\\
  Atlanta GA 30332 \\
  \texttt{minah.lee@gatech.edu} \\
}

\begin{document}
\maketitle

\begin{abstract}

This paper proposes a novel problem: vision-based perception to learn and predict the collective dynamics of multi-agent systems, specifically focusing on interaction strength and convergence time. Multi-agent systems are defined as collections of more than ten interacting agents that exhibit complex group behaviors. Unlike prior studies that assume knowledge of agent positions, we focus on deep learning models to \textbf{directly predict collective dynamics from visual data}, captured as frames or events. Due to the lack of relevant datasets, we create a simulated dataset using a state-of-the-art flocking simulator, coupled with a vision-to-event conversion framework. We empirically demonstrate the effectiveness of event-based representation over traditional frame-based methods in predicting these collective behaviors. Based on our analysis, we present \textbf{e}vent-based \textbf{v}ision for \textbf{M}ulti-\textbf{A}gent dynamic \textbf{P}rediction (\textbf{evMAP}), a deep learning architecture designed for real-time, accurate understanding of interaction strength and collective behavior emergence in multi-agent systems.

\end{abstract}
\section{Introduction}
\label{sec:intro}

\begin{figure}[tb]
  \centering
   % \vspace{-0.1in}
   \includegraphics[width=0.7\linewidth]{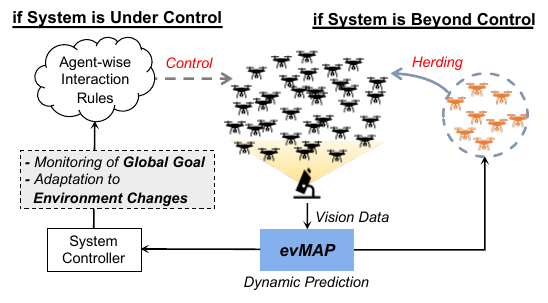}
   % \vspace{-0.2in}
   \caption{\textbf{ Application examples of collective dynamic prediction of multi-agent system.} Multi-agent dynamic prediction is helpful for both systems that are under and beyond control.}
   \label{fig:motivation}
   % \vspace{-0.2in}
\end{figure}

The systems of large number ($>$10) of agents, hereafter referred to as a multi-agent system, are crucial in a wide range of autonomy applications, including swarm robotics~\cite{Khamis2015MultiRobot} and fleets of autonomous vehicles~\cite{RiosTorres2017Coordination}. Inspired by collective behaviors observed in nature such as fish schools and bird flocks, these systems aim to achieve collective goals through the interaction among individual agents using a set of decentralized rules. Analytical flocking models such as Reynolds model~\cite{reynolds1987flocks} or Vicsek model~\cite{vicsek1995novel} replicate collective behaviors observed in nature, but these models require precise localization which is rarely possible in the real-world applications. Therefore, real-time prediction of collective behavior,  like how and when agents will achieve a collective goal, is essential for adapting the local rules and controlling multi-agent systems in a real-world environment~\cite{viragh2014flocking, shrit2021i2sl} as illustrated in Figure~\ref{fig:motivation}. This prediction is valuable in competitive settings like swarm herding~\cite{chipade2020multi}, where understanding the system dynamics of adversarial agents can enhance strategic control. The prediction is crucial for optimizing resources and minimizing risks in complex operations, such as coordinating astrobots in telescopes~\cite{macktoobian2021learning, macktoobian2022data}, where precise maneuvering and dense formations are important. As swarm operations scale in complexity, the prediction of collective behavior becomes increasingly critical, underscoring the need for advancement in methods for learning and control of multi-agent systems.

\begin{figure}[tb]
  \centering
   % % \vspace{-0.1in}
   \includegraphics[width=0.6\linewidth]{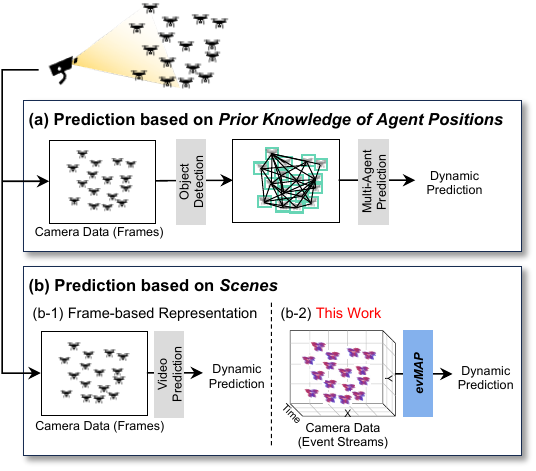}
   % \vspace{-0.1in}
   \caption{\textbf{Several methods for understanding dynamics in a multi-agent system.} (a) Many previous studies in multi-agent prediction require pre-processing for detecting agents. This paper focuses on scene-based perception: compared to (b-1) frame-based methods, (b-2) event-based methods demonstrate their effectiveness in understanding multi-agent dynamics.}
   \label{fig:intro}
   \vspace{-0.2in}
\end{figure}

This paper introduces the novel problem of \textbf{\textit{real-time prediction of collective behavior in multi-agent dynamics from visual observations}}. Many previous studies on multi-agent systems assume \textit{prior knowledge of agent states} and are primarily designed to predict individual agent trajectories~\cite{vishnu2023improving, luo2023jfp, mehr2023maximum}. A potential approach involves integrating object detection with trajectory predictors (Figure~\ref{fig:intro}(a)). However, challenges exist in both object detection and trajectory prediction for understanding multi-agent dynamics. 

\begin{figure}[tb]
  \centering
   % \vspace{-0.1in}
   \includegraphics[width=0.8\linewidth]{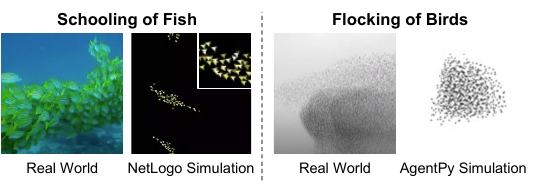}
   % \vspace{-0.1in}
   \caption{\textbf{Sample flocking scenes~\cite{yt_school, yt_flock} and simulators (NetLogo~\cite{netlogoflocking}, AgentPy~\cite{foramitti2021agentpy}). }}
   \label{fig:yt_sample}
   \vspace{-0.2in}
\end{figure}

In object detection, the small size of agents and their high density in the scenes (Figure~\ref{fig:yt_sample}) hinder deep learning models from accurately determining agent positions~\cite{yuan2023small, su2023opa}. For trajectory prediction, inferring agent-wise trajectories and collective behavior from multiple trajectories becomes computationally infeasible with $M \times T \times N$ trajectories ($M$: number of agents, $T$: sequence length, $N$: number of sequences). Moreover, combining a multi-agent trajectory predictor with an object detector requires substantial computational resources. Even with exact positions provided using GPS augmented for each agent (e.g., drones) instead of object detection, high energy usage and latency from continuous GPS usage, as well as the computational burden from multi-agent trajectory predictions, present challenges in employing state (position)-based multi-agent prediction models for systems with large number of agents.

In contrast, we are inspired by the success of deep learning (DL) methods in learning dynamics for prediction and control from visual inputs without the need for exact state knowledge~\cite{ze2023visual, yarats2021improving, yarats2021image, yarats2021reinforcement}. While existing vision-to-dynamics models have been demonstrated for systems with a few agents, predicting the dynamics of collective multi-agent systems (with more than 10 agents) from vision remains an unexplored area. Our objective is to \textbf{\textit{directly learn and predict the collective dynamics (not the states of each agent) of a multi-agent system from scenes}} captured via frame and event-based cameras, as illustrated in Figure~\ref{fig:intro}(b).

We propose leveraging advancements in event cameras~\cite{gallego2020event}, which capture per-pixel brightness changes with high temporal resolution and dynamic range, to predict multi-agent collective dynamics using vision. Event-based vision has recently achieved significant improvements in object recognition, detection, and segmentation, as well as tracking high-speed objects~\cite{gehrig2023recurrent, peng2023get, hamaguchi2023hierarchical, li2022asynchronous, he2021fast}. However, applying it to understand multi-agent dynamics remains largely unexplored (Figure~\ref{fig:intro}(b-2)).

To address the lack of datasets for vision-based analysis of large interacting agent groups, we have created a new dataset based on Reynolds' rule~\cite{reynolds1987flocks} using NetLogo~\cite{netlogoflocking} and AgentPy~\cite{foramitti2021agentpy}. Frame-based inputs were generated (Figure~\ref{fig:yt_sample}) and converted to event-based data using the v2e~\cite{hu2021v2e} framework. Our results demonstrate that event-based methods outperform frame-based methods in capturing real-time dynamics and predicting collective behavior from early observations. Additionally, our proposed model shows superior performance in capturing time-varying dynamics compared to other event-based approaches.

This paper makes following unique contributions:
\begin{itemize}
    \item This paper introduces a novel problem, \textbf{\textit{vision to prediction of collective multi-agent dynamics}} for real-time perception and control of multi-agent system. We study deep learning models for prediction of collective  multi-agent dynamics from frame- and event- based visual inputs. To the best of our knowledge, this is the first study to discuss multi-agent dynamic prediction from visual observation.
    \item This paper performs a comparative study between frame- and event-based methods, and empirically demonstrate the advantage of event representation in learning and predicting collective behavior of multi-agent systems. Prior works have studied processing event representation for various tasks, but to the best of our knowledge, this is the first work demonstrating event-based methods for predicting multi-agent dynamics. 
    \item We present a new transformer-based deep learning architecture, \textbf{\textit{e}}vent-based \textbf{\textit{v}}ision for \textbf{M}ulti-\textbf{A}gent dynamic \textbf{P}rediction (\textbf{evMAP}), to learn dynamics of multi-agent systems. In particular, the model is designed to efficiently recognize dynamic changes in the multi-agent systems.  

\end{itemize}
          % 3p
\section{Related Works}
\label{sec:related}

Multi-agent prediction has been widely studied in fields like robotics and autonomous driving. In particular, trajectory forecasting in multi-agent system often incorporates memory mechanisms to model agent interactions and temporal dependencies~\cite{xu2022remember}. While our approach shares similarities with these methods, focusing on agent interactions, it differs in two ways: 1) using vision data instead of agent states (positions), and 2) predicting collective behavior at the system level rather than individual trajectories. Additionally, event-based vision, known for its ability to capture high-temporal resolution changes in dynamic environments~\cite{gallego2020event}, has been successfully applied to tasks such as object detection and tracking~\cite{gehrig2023recurrent}. However, its application to multi-agent system dynamics remains underexplored, which we address by using event-based data to predict collective behaviors in real time.
\section{Vision to Multi-Agent Dynamic Prediction}

\subsection{Multi-Agent Simulation and Visual Data Preparation.}
\label{sec:data_gen}

\begin{figure}[tb]
  \centering
  % \vspace{-0.1in}
   \includegraphics[width=0.7\linewidth]{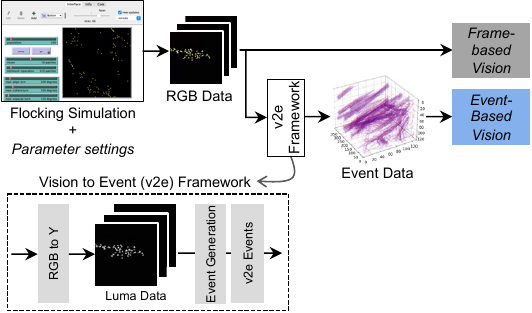}
    % \vspace{-0.1in}
   \caption{ \textbf{Simulation framework of frame- and event-based vision for multi-agent dynamic prediction.} Due to the absence of existing dataset, flocking simulations~\cite{netlogoflocking, foramitti2021agentpy} and event synthesis toolbox~\cite{hu2021v2e} are used to generate multi-agent dynamic sequence and convert from frame to event.}
   \label{fig:sim_framework}
   % \vspace{-0.2in}
\end{figure}

Due to the scarcity of publicly available vision datasets that capture the collective behavior of more than ten agents based on their interactions, we employ flocking simulators (NetLogo~\cite{netlogoflocking}, AgentPy~\cite{foramitti2021agentpy}) and the event synthesis framework~\cite{hu2021v2e} to create an event-based dataset for evMAP (Figure~\ref{fig:sim_framework}). Craig W. Reynolds proposed three fundamental rules for realistic modeling of flocking behaviors~\cite{reynolds1987flocks}. Using these rules, we generate frames from the simulators that capture various collective behaviors emerging from multi-agent interactions. These frames are used for training and evaluating frame-based prediction models. Furthermore, the frames are converted into event data using the v2e toolbox~\cite{hu2021v2e}, which synthesizes realistic event data from intensity frames while accounting for non-idealities in event cameras, such as noise and motion blur. This event data is then used to train and evaluate event-based pmodels. The resulting dataset consists of several sub-datasets, each simulating different levels of agent interaction strengths to reflect varying interaction intensities and convergence times (see Supplementary for details).

\begin{figure}[tb]
  \centering
  % \vspace{-0.1in}
   \includegraphics[width=0.8\linewidth]{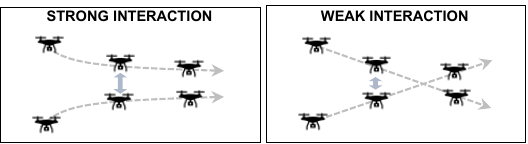}
   % \vspace{-0.1in}
   \caption{ \textbf{(left) Strong interactions cause large changes in agents' velocities and (right) weak interactions relatively maintain individual agents' velocities.}}
   \label{fig:interactions}
   % \vspace{-0.1in}
\end{figure}

\subsection{Prediction Target and Evaluation Metric}
\label{sec:prediction_target}

\begin{figure}[tb]
  \centering
  % \vspace{-0.05in}
   \includegraphics[width=\linewidth]{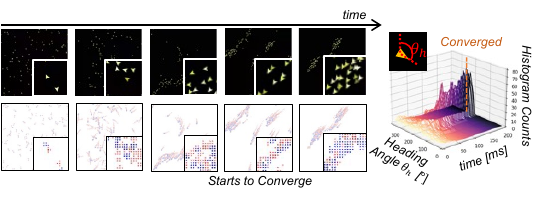}
   % \vspace{-0.35in}
   \caption{ \textbf{Multi-agent swarming behavior in frame (top row) and event (bottom row) domains.} Events are aggregated into frames for every 1ms for visualization purposes. The heading directions of all agents converge to a certain direction from a specific time, referred to as \textit{convergence-time}.}
   \label{fig:swarm_sample}
   % \vspace{-0.2in}
\end{figure}

In this paper, we focus on learning and predicting collective behavior of agents. First, we consider the task of predicting \textbf{interaction strengths between agents} in Figure~\ref{fig:interactions}, which is critical for learning (or imitating) collective dynamics of swarming agents. Along with "static" interaction strength, we define the task to also recognize dynamic changes in the interaction strength, which is important to understand changes in the swarm intents.

Second task is predicting the \textbf{convergence time of the swarms}, an outcome of the collective behavior. For example, collective behavior of 100 agents under flocking rules is demonstrated using NetLogo simulation~\cite{netlogoflocking} in Figure~\ref{fig:swarm_sample}. Initially, each agent, represented as a yellow arrow, starts at a random position and direction. They interact based on flocking rules and eventually gathering and aligning their directions. Convergence occurs when most agents move in the same direction. We mathematically define \textit{convergence-time} as the moment when the majority of agents head in a unified direction (see Supplementary for details).

\begin{figure}[tb]
  \centering
  % \vspace{-0.2in}
   \includegraphics[width=\linewidth]{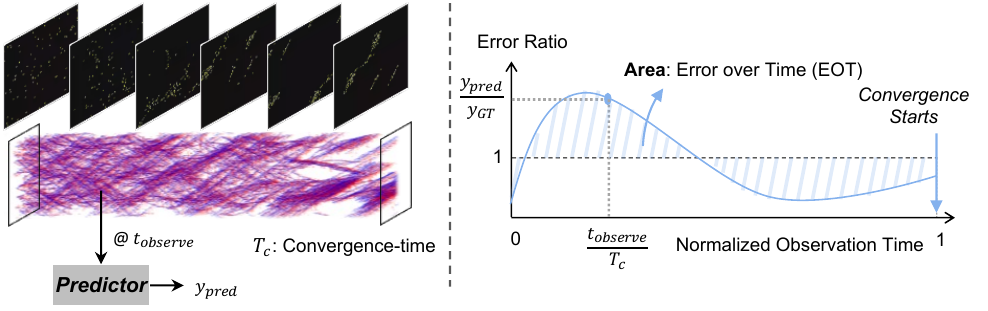}
    % \vspace{-0.1in}
   \caption{\textbf{(top) Multi-agent dynamic prediction over time in frame sequence or event stream, (bottom) illustration of error ratio and EOT} as quantitative measures for predicting (interaction strength, convergence-time).}
   \label{fig:auec}
   % \vspace{-0.2in}
\end{figure}

We have developed an \textbf{evaluation metric} to quantitatively assess the multi-agent interaction predictioins in a sequence. As described in Figure~\ref{fig:auec}, the predictor provides continuous estimations (interaction strength or convergence-time), and the prediction $y_{pred}$ is compared to the ground truth $y_{GT}$ using an error ratio $y_{pred}$/$y_{GT}$. The ratio is chosen due to significant variation in ground truth values across sequences. The error curve plots this ratio over normalized observation time scale $t_{observe}$/$T_{c}$, with deviations from 1 indicating inaccuracy. The area under the curve, referred to as \textit{error over time} (EOT), quantifies the model's accuracy, with lower EOT reflecting better temporal prediction quality.

\subsection{Problem Formulation}

Vision-based multi-agent dynamic prediction aims to forecast the collective dynamics of a multi-agent system with a large number ($>$ 10) of interacting agents. Assume that a collection of $M (>10)$ interacting agents forms a collective dynamic system. An event-based camera captures a set of events \(\mathcal{E}_\tau = \{ (x_i, y_i, t_i, p_i) \mid t_i \in \tau, T = \sup t_i \}\) during a time interval \(\tau\), where \((x_i, y_i)\) (\(0 \leq x_i \leq W, 0 \leq y_i \leq H\)) denotes pixel locations, \(t_i\) represents event triggering timestamps, \(p_i \in \{-1, 1\}\) indicates the polarity (relative change in brightness), and \(T\) is the total observation time. Our goal is to predict the collective dynamics \(\mathcal{D}_{\tau} \in \mathbb{R}^n\) of the multi-agent system. Specifically, agent-wise interaction strength (Section \ref{sec:exp_interaction_strength}) and collective behavior emergence time (Section \ref{sec:exp_time}) are considered as examples of \(\mathcal{D}_{\tau} \in \mathbb{R}\).

Note that we consider each agent with a size (not as a point), generating multiple events. Our goal is to directly predict collective behavior from event camera data, without mapping events to individual agents.

\subsection{Prediction Models}

The concept of vision to multi-agent dynamic prediction is new, and there is no existing prediction models designed specifically for this task. We consider prior video recognition models for frame-based prediction and event-based object recognition models for event-based prediction. 

\paragraph{\textbf{Frame-based Models.}}

We examine SlowFast~\cite{feichtenhofer2019slowfast}, which uses a dual-pathway approach combining a slow pathway for spatial semantics and a fast pathway for motion. We also investigate frame-based transformer model, MoViTv1~\cite{fan2021multiscale}, which a multi-scale pyramid of features, capturing simple visual information in early layers and complex, high-dimensional features in deeper layers. MoViTv2~\cite{li2022mvitv2}, an enhancement of MoViTv1, integrates decomposed relative positional embedding and residual pooling connections to maintain shift-variance. 

\paragraph{\textbf{Event-based Models.}}
We assess AEGNN~\cite{schaefer2022aegnn}, an event graph-based method that limits re-computation, thereby significantly reducing computation time and latency. Additionally, we examine Eventformer~\cite{kamal2023associative}, a transformer-based approach that leverages an associative memory mechanism for efficient event processing. 
For all of these models, we replace their classification heads with regression heads, retraining them with L1 loss for predicting interaction intensity and convergence-time (see Supplementary for detail).

\begin{figure}[tb]
  \centering
  \vspace{-0.3in}
   \includegraphics[width=0.5\linewidth]{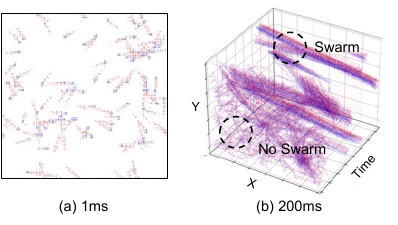}
    \vspace{-0.15in}
   \caption{\textbf{Event data from Multi-Agent Dynamics;} (a) Event-based frame from 1ms aggregation, (b) Event-based stream from 200ms aggregation}
   \vspace{-0.2in}
   \label{fig:swarm_event}
   % \vspace{-0.1in}
\end{figure}

\paragraph{\textbf{Unique Challenges for Multi-Agent Dynamic Prediction.}}
The prediction of collective multi-agent dynamics from events has unique challenges compared to prior event-based tasks such as semantic segmentation~\cite{alonso2019evsegnet, sun22ess}, object detection~\cite{gehrig2023recurrent}. We argue that, while spatiotemporal information is useful, these tasks can achieve good accuracy using solely spatial information~\cite{yang2023event}. For example, analyzing an event-based frame for a brief time span allows for the segmentation of road regions, detection of cars, or depth estimation of different areas. However, predicting agent interactions or collective behavior becomes significantly easier with the observation over extended periods (Figure~\ref{fig:swarm_event}), highlighting the importance of long-term spatiotemporal information for understanding multi-agent dynamics. Additionally, effectively using short-term and long-term observation is crucial for adapting to changes in system dynamics. In the following section, we present a new event-based model, evMAP, specially designed model to address these challenges for multi-agent dynamic prediction.
        % 1p
\section{\textbf{\textit{e}}vent-based \textbf{\textit{v}}ision for \textbf{M}ulti-\textbf{A}gent dynamic \textbf{P}rediction}

\begin{figure}[tb]
  \centering
  % \vspace{-0.2in}
   \includegraphics[width=0.8\linewidth]{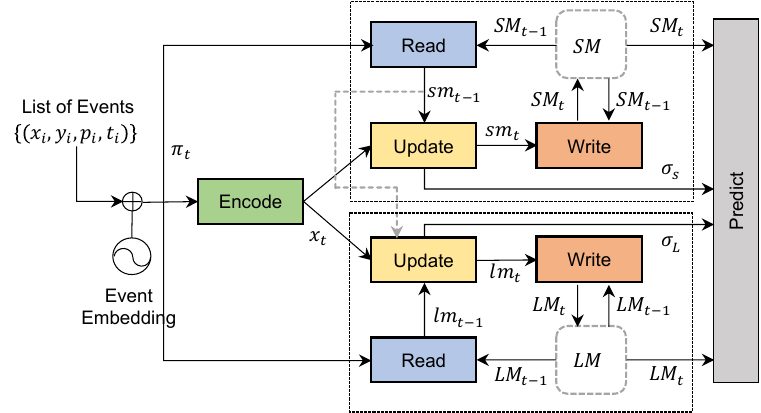}
    % \vspace{-0.3in}
   \caption{\textbf{evMAP Architecture.} evMAP contains two memories to effectively store recent spatial and long-term spatiotemporal observations, and adaptively exploit them for fast, accurate prediction under dynamic changes.}
   \label{fig:evmap}
   % \vspace{-0.2in}
\end{figure}

This section introduces evMAP, ML model specially designed for understanding collective behavior of multi-agent system. As described in Figure~\ref{fig:evmap}, evMAP uses event embedding $\pi_t$ and an encoder to calculate pair-wise spatiotemporal interactions $x_t$ among events. $x_t$ along with previous observations $sm_{t-1}, lm_{t-1}$ retrieved from associative memories, are used to generate updated spatiotemporal information $sm_t, lm_t$. Then, short-term memory $SM$ (focusing on recent dynamics) and long-term memory $LM$ (covering broader past behavior patterns) are updated based on these information.  The updated memory contents $SM_t, LM_t$ are then merged according to their update amounts $\sigma_S, \sigma_L$ for use in prediction tasks through a classification or regression head.

\begin{figure}[tb]
  \centering
  % \vspace{-0.2in}
   \includegraphics[width=\linewidth]{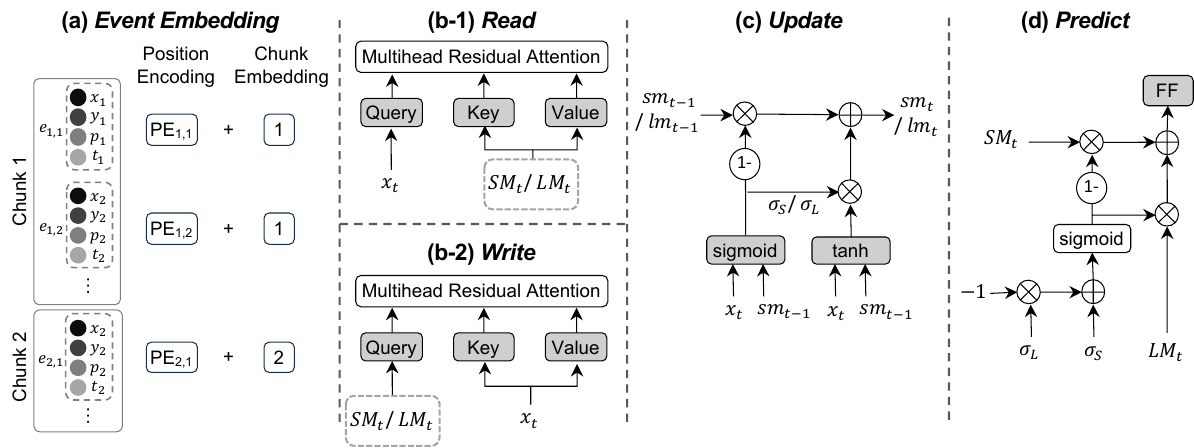}
    % \vspace{-0.3in}
   \caption{\textbf{Computation Blocks.} \texttt{Read}, \texttt{Write}, \texttt{Update} are processed separately for short-term and long-term memories. Grey colored blocks include a fully connect layer.}
   \label{fig:evmap_detail}
   % \vspace{-0.2in}
\end{figure}

\paragraph{\textbf{Event Embedding:}}
An event stream is chunked by a certain time length and evMAP is processing on each chunk. An event stream is converted into normalized position coordinates (2D), polarity (1D), normalized time within a chunk (1D), and chunk embedding (1D). This low-dimensional (5D) data is mapped into a higher dimensional feature space using a learnable Fourier feature-based positional encoder (Figure~\ref{fig:evmap_detail}(a)). 

\paragraph{\textbf{Encode:}}
Self-attention among the positional embedding $\pi_t$ is employed to extract pair-wise interactions within an event chunk. Refine algorithm from \cite{kamal2023associative} is adopted for an efficient computations.

\paragraph{\textbf{Read \& Write:}}
\texttt{Read} operation employs cross-attention between the current observation $\pi_t$ and past memory contents to retrieve past memories $M_{t-1}$ ($SM_{t-1}$ or $LM_{t-1}$) relevant to $\pi_t$ (Figure~\ref{fig:evmap_detail}(b-1)). During \texttt{Write}, cross-attention is applied between $M_{t-1}$ and current spatiotemporal information $m_t$ ($sm_t$ or $lm_t$)(Figure~\ref{fig:evmap_detail}(b-2)). While \texttt{Read} and \texttt{Write} modules are adopted from \cite{kamal2023associative}, \texttt{Erase} operation following \texttt{Write} is omitted, as the removal of past memories is managed during \texttt{Update}.

\paragraph{\textbf{Update:}}
New spatiotemporal information $m'_t$ ($sm'_t$ or $lm'_t$) and update gate $\sigma_m$ ($\sigma_S$ or $\sigma_L$) are each computed based on the current encoded observation $x_t$ and $sm_{t-1}$, and updated based on the past information ($sm_{t-1}$ or $lm_{t-1}$) as follows:
% \vspace{-0.1in}
\begin{gather}
m'_t = \tanh(W_{xm} x_t, W_{mm}sm_{t-1})\\
\sigma_m = \sigma (W_{x} x_t + W_{m}sm_{t-1})\\
m_t = (1 - \sigma_m) \odot m_{t-1} + \sigma_m \odot m'_t
\end{gather}
where $\odot$ denotes element-wise multiplication (Figure~\ref{fig:evmap_detail}(c)). Small $\sigma_m$ value corresponds to less new information to be updated, whereas a high $\sigma_m$ value suggests more new information. \texttt{Update} process is executed separately on both short-term $sm_t$ and long-term memories $lm_t$.

\paragraph{\textbf{Predict:}}
Short-term memory ($SM_t$) captures recent individual agent dynamics, while long-term memory ($LM_t$) stores broader past information related to group behavior. Both are crucial for predicting future collective behavior, but their importance shifts with system dynamics. In weak interactions, where individual dynamics are simpler, $LM_t$ becomes more relevant, while in strong interactions, the complexity of individual dynamics requires a greater focus on $SM_t$. An adaptive strategy balances the two based on system dynamics, with $\sigma_S$ and $\sigma_L$ indicating the emphasis on $SM_t$ or $LM_t$, respectively. The final prediction is made by weighting contributions from both memories as follows:
% \vspace{-0.3in}
\begin{gather*}
\sigma_{SL} = \sigma(\sigma_S - \sigma_L) \\
y' = FC (\sigma_{SL} \odot SM_t + (1 - \sigma_{SL}) \odot LM_t)
\end{gather*}

\paragraph{\textbf{Comparison to Frame-based Models.}}

Traditional video recognition models such as SlowFast~\cite{feichtenhofer2019slowfast}, MoViTv1~\cite{fan2021multiscale}, and MoViTv2~\cite{li2022mvitv2} can be considered for vision-based perception in understanding the collective behaviors of multi-agent systems. These models process sequences of images that store the absolute brightness of every pixel. While images contain rich spatial information, such as shape, color, and texture, this information is not very helpful for understanding multi-agent systems with a large number of agents, where each agent occupies fewer than 50 pixels and forms very dense groups (Figure~\ref{fig:yt_sample}). Additionally, a large area of each image is occupied by the background due to the small size of the agents, making it challenging for the model to focus on the agent dynamics. In contrast, evMAP processes event streams that inherently store only the agent dynamics with high temporal resolution, making it easier for the model to focus on the agents.

\paragraph{\textbf{Comparison to Eventformer.}}
Eventformer~\cite{kamal2023associative} incorporates GRU-based computation with associative memory that reads and writes information linked to GRU's hidden states, emphasizing spatiotemporal features. However, its update mechanism applies gating operations twice—once in the GRU and again in the erase computation—potentially leading to an overload of past information. This is helpful for object recognition tasks in static scenes as well as multi-agent dynamic predictions with fixed dynamic rules, but not for multi-agent system where its dynamics change. In contrast, evMAP uses two separate associative memories to store near-past \textbf{\textit{spatio}}temporal and long-range spatio\textbf{\textit{temporal}} information, ensuring the preservation of information from two different perspectives. Moreover, \texttt{Predict} module in evMAP facilitates the adjustments to dynamic system changes, in which relative experiments demonstrated in Section~\ref{sec:dynamic_change}.

           % 2p
\section{Simulation Results}

\subsection{Agent-wise Interaction Strength Detection}
\label{sec:exp_interaction_strength}

\begin{figure}[tb]
  \centering
   % \vspace{-0.1in}
   \includegraphics[width=\linewidth]{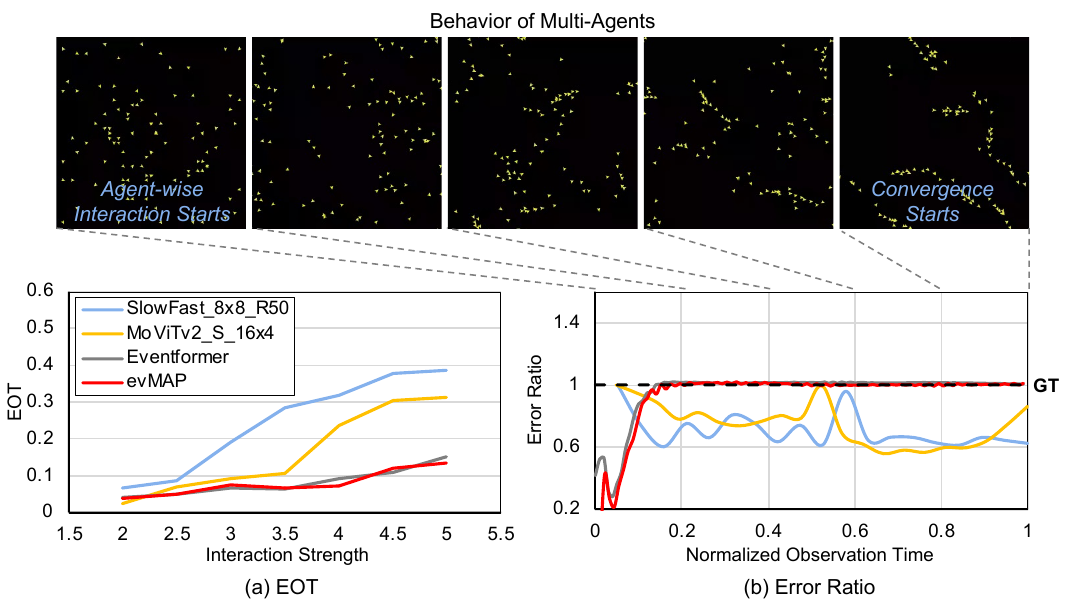}
    % \vspace{-0.3in}
   \caption{(a) \textbf{EOT of interaction strength prediction} across weak to strong interactions, (b) \textbf{Interaction strength prediction errors} from various frame- and event-based models (interaction strength=3). Definitions of error ratio, EOT, normalized observation time are discussed in Section~\ref{sec:prediction_target}. }
   % \vspace{-0.2in}
   \label{fig:error_turn}
\end{figure}

Figure~\ref{fig:error_turn}(a) shows the EOT (discussed in Section~\ref{sec:prediction_target}) for interaction strength detection. Both frame-based (SlowFast, MoViTv2) and event-based (Eventformer, evMAP) methods demonstrate comparable quality in detecting weak interactions, while event-based processing excels in identifying stronger interactions. The effectiveness of \textit{\textbf{event representation}} in detecting interaction strength primarily stems from its high temporal resolution, resulting more fine-grained temporal data. During the same time interval, event data can reveal the direction and speed of each agent's heading, which frames cannot capture, as shown in Figure~\ref{fig:swarm_sample}. Temporal information is crucial for detecting interaction strength, making event representations more effective.

\paragraph{\textbf{Ability of Early Prediction.}}

Figure~\ref{fig:error_turn}(b) presents the prediction results across an observation time sequence, from the beginning of agent interactions to convergence. 
Note that frame-based models require a specific observation window, leading to their initial predictions only becoming available after a certain period from the start of the observations. This delay is particularly significant when detecting stronger interactions, as the convergence time decreases while the observation window size remain fixed, resulting in a relatively longer wait before the initial prediction can be made. 

Frame-based methods (SlowFast, MoViTv2) tend to provide predictions that are higher than the actual interaction strength under weak interaction and lower predictions under stronger interactions, indicating their difficulty to differentiate between varying interaction strengths. In contrast, the event-based models (Eventformer, evMAP) consistently provide accurate predictions (error ratio between 0.95 to 1.05) from an early prediction stage (before 25\% of the pre-convergence interaction time) across all levels of interaction strengths (see Appendix for detail).

\subsection{Emergent Time Prediction of Collective Behavior} 
\label{sec:exp_time}

\begin{figure}[tb]
  \centering
  % \vspace{-0.1in}
   \includegraphics[width=\linewidth]{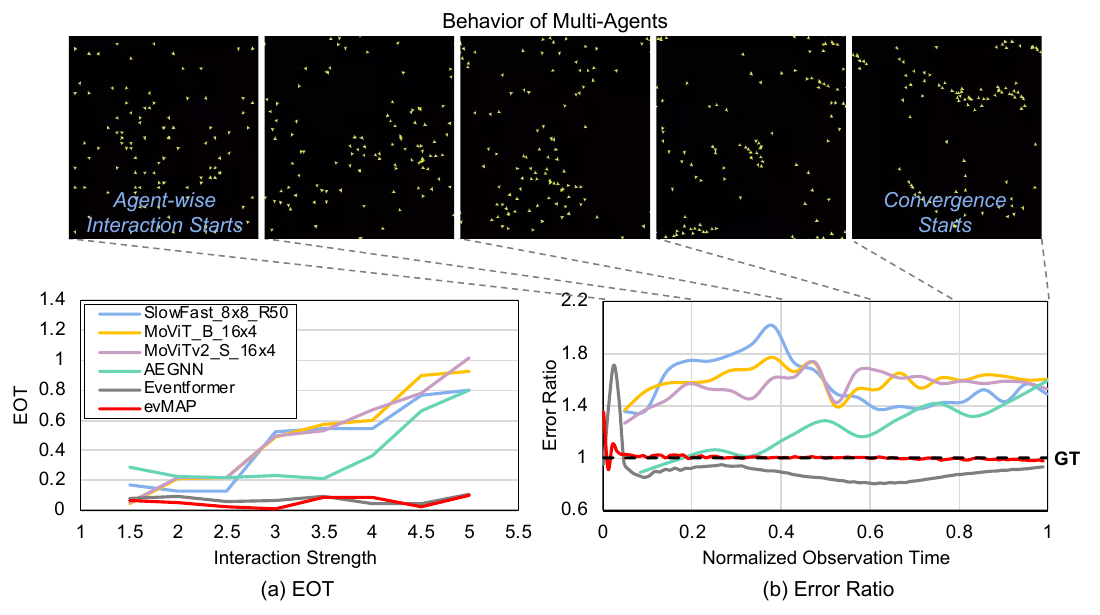}
    % \vspace{-0.3in}
   \caption{(a) \textbf{EOT for convergence-time prediction} across weak to strong interactions. (b) \textbf{Convergence-time prediction errors} from various frame- and event-based models (interaction strength=3). Definitions of error ratio, EOT, normalized observation time are discussed in Section~\ref{sec:prediction_target}.}
   \label{fig:time_error}
   % \vspace{-0.2in}
\end{figure}

We evaluate the convergence-time prediction of various frame- and event-based models under different interaction strengths, as shown in Figure~\ref{fig:time_error}(a). All frame-based methods (SlowFast, MoViTv1, MoViTv2) exhibit high-quality predictions (EOT $<$ 0.5) under weak agent-wise interactions. However, their error increases substantially under strong interactions, suggesting that frame-based approaches struggle with understanding complex agent dynamics. In contrast, an event graph-based method (AEGNN) demonstrates high-quality predictions for weak and intermediate strength of agent interactions. \textit{\textbf{Event representation}} plays a critical role in accurately understanding the interactions among agents and their collective behavior, which aids in more accurate convergence-time predictions. Nonetheless, it struggles with prediction under stronger interactions. This limitation primarily arises because computations based on event graphs are resource-intensive, necessitating a reduction in the number of processed events. This reduction under strong agent-wise interactions leads to the loss of non-linear behaviors of agents, which in turn affects accuracy. 

On the other hand, the event-based methods using transformer architecture (Eventformer, evMAP) excel in making highly accurate predictions across all levels of interaction strengths. This enhanced performance is attributed to its ability to \textit{\textbf{process the entire set}} of event data without the need for event reduction for computational efficiency. This advantage is particularly valuable under strong interactions among agents, where agents exhibit abrupt behaviors.

\paragraph{\textbf{Ability of Early Prediction.}}
Figure~\ref{fig:time_error}(b) presents the prediction results across an observation time sequence. 
Under weak interactions, all frame-based methods (SlowFast, MoViTv1, MoViTv2) demonstrate high-quality predictions (error ratio between 0.95 to 1.05) from an early prediction stage (before 20\% of the pre-convergence interaction time), due to the relatively simple and predictable interactions among agents. However, under stronger interactions, the prediction quality significantly decreases. Meanwhile, an event-based method using a graph neural network (AEGNN) also encounters difficulties with strong interactions as well as early stage of predictions.

In contrast, the event-based methods using transformer architecture (Eventformer, evMAP) achieve highly accurate predictions from an early stage (before 10\% of the pre-convergence interaction time) across all strengths of interactions. Its \textit{\textbf{associative memory}} helps efficient accumulation and utilization of knowledge from long-term observations, enabling early and accurate predictions. Notably, our evMAP excels in providing very early and correct predictions, particularly for weak and intermediate interactions, by focusing more on spatial dynamics for prediction while effectively accumulating observations in long-term memory. Although its error ratio for strong interactions right before convergence appears significant, considering the absolute convergence time for strong interactions is brief, the absolute error is minor (see Appendix for detail).

In summary, the event representation, which entails processing the entire data and employing associative memory, appears advantageous for understanding both the interaction strength among agents (agent-wise dynamics) and the convergence time (collective behavior) of multi-agent systems in real-time.

\subsection{System Dynamic Change Detection} 
\label{sec:dynamic_change}

\begin{figure}[tb]
  \centering
  % \vspace{-0.1in}
   \includegraphics[width=\linewidth]{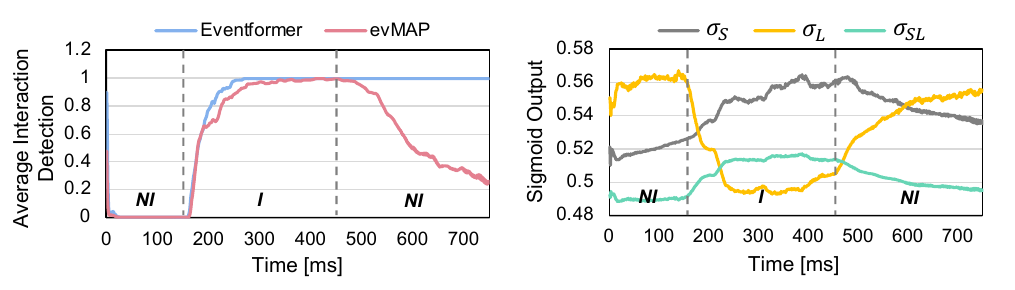}
    % \vspace{-0.1in}
   \caption{\textbf{(top) Agent-wise interaction detection of Eventformer~\cite{kamal2023associative} and evMAP.} Detection 0 indicates absence of interaction and 1 indicates the presence of interaction. \textbf{(bottom) Updates and adaptation of short-term and long-term memory in evMAP.} \textit{NI} denotes system without agent-wise interaction and \textit{I} denote system with interaction.}
   \label{fig:dynamic_change}
   % \vspace{-0.2in}
\end{figure}

Previous experiments show that processing event representations through associative memory can facilitate an early and accurate understanding of multi-agent system dynamics. However, a significant limitation of using associative memory is its inability to quickly adapt to sudden changes in system dynamics, such as abrupt climate shifts or external control interventions, as its prediction relies on the entire past observation.

For two models based on associative memory, Eventformer~\cite{kamal2023associative} and evMAP, we replace their regression head to classification head, and train with cross-entropy loss to differentiate between situations with and without multi-agent interactions. Subsequently, we evaluate their capabilities of recognizing agent-wise interactions when it emerges and then vanishes. As illustrated in Figure~\ref{fig:dynamic_change}, upon the appearance of agent-wise interactions at 150ms, both models quickly ($\sim$ 15ms) identify the presence of interactions. In evMAP, $\sigma_L$ (amount of updates in long-term memory) decreases with the onset of agent-wise interaction, as agents begin exhibiting collective behavior with stronger spatiotemporal patterns, indicating fewer new information updates are needed. On the other hand, $\sigma_S$ (amount of updates in short-term memory) increases, suggesting the final prediction of evMAP shifts its focus towards spatial information. When interactions disappear at 450ms, Eventformer fails to recognize this change, continuing to predict ongoing interactions. In contrast, evMAP accurately identifies the shift, adjusting its focus towards temporal dynamics in response to the reduced spatiotemporal correlations, thereby learning from the new dynamics. Its detection confidence is not as high as the initial detection however, as evMAP's focus on temporal dynamic is hindered by its past observation of system with interactions.

\begin{table*}[tb]
% \vspace{-0.1in}
  \caption{\textbf{Performance comparison among several frame- and event-based methods.} For frame-based methods, an input image size of 224$\times$224 is considered. Each sequence contains, on average, 200 frames or 550K events.}
   % \vspace{-0.1in}
  \label{tab:complexity}
  % \resizebox{\textwidth}{!}{%
  \centering
  \setlength{\tabcolsep}{6pt}
  \begin{tabular}{@{}lcccccc@{}}
    \toprule
    Model &Represent.& Architecture & \multicolumn{1}{c}{\begin{tabular}[c]{@{}c@{}} Avg. EOT $\downarrow$\\ (Interaction\\ Strengths) \end{tabular}} & \multicolumn{1}{c}{\begin{tabular}[c]{@{}c@{}}Avg. EOT $\downarrow$\\ (Convergence\\ Time)\end{tabular}} & \multicolumn{1}{c}{\begin{tabular}[c]{@{}c@{}}Params\\ {[}MB{]} $\downarrow$\end{tabular}} & \multicolumn{1}{c}{\begin{tabular}[c]{@{}c@{}}TFLOPS\\ /sequence $\downarrow$\end{tabular}}\\
    \midrule
    SlowFast~\cite{feichtenhofer2019slowfast} & Image Frame & CNN &0.2453& 0.491 & 53.0 & 27.5\\
    % SlowFast (R)~\cite{feichtenhofer2019slowfast} & RGB Frame, Convolution & 1.077 & - & 53.0 & 5.51\\
    MoViT~\cite{fan2021multiscale} &Image Frame& Transformer &-& 0.559 & 36.4 & 14.1\\
    % MoViTv1 (R)~\cite{fan2021multiscale} & RGB Frame, Transformer & 0.889 & - & 36.4 & 2.82\\
    MoViTv2~\cite{li2022mvitv2} &Image Frame&Transformer&0.1641& 0.562 & 34.3 & 12.9\\
    % MoViTv2 (R)~\cite{li2022mvitv2} & RGB Frame, Transformer & 1.155 & - & 34.3 & 2.58\\
    \midrule
     % & Event, Point Cloud &&&&\\
    AEGNN~\cite{schaefer2022aegnn}&Event Graph& Graph & - & 0.390 & 0.0298 & 0.190\\
    \midrule
    Eventformer~\cite{kamal2023associative} & Event Set& Transformer & \underline{0.0825} & \underline{0.074} & \textbf{0.011} & \textbf{0.0192}\\
    \textbf{evMAP} &Event Set& Transformer & \textbf{0.0796} & \textbf{0.055} & \underline{0.013} & \underline{0.0216} \\
  \bottomrule
   % \vspace{-0.3in}
\end{tabular}
% }
\end{table*}

\paragraph{\textbf{Computation Complexity.}}
Table~\ref{tab:complexity} compares the computational overheads of different models for understanding multi-agent dynamics, presenting total FLOPS per sequence for direct comparison. Frame-based models (SlowFast, MoViTv1, MoViTv2) incur higher computational costs, while event-based methods (AEGNN, Eventformer, evMAP) process events more efficiently, reducing computation costs and enabling real-time processing. Notably, Eventformer and evMAP handle unstructured event streams without requiring event-graph formulation, resulting in the lowest FLOPS.

\paragraph{\textbf{Simulated Data and Real-World Applicability.}}
Due to the high complexity of capturing real-world swarms of more than ten agents using both visible and event cameras, 
the experiments presented in this work rely on simulated swarm data. Our dataset is generated based on well-established interaction rules, which is intended to capture the fundamental properties of real-world swarm behavior~\cite{reynolds1987flocks}. Moreover, the generated event-based data also accounts for noise and non-idealities, reflecting real-world conditions~\cite{hu2021v2e}. We have also evaluated evMAP on simulated data that imitates real-world scenarios, as detailed in the Supplementary.             % 6p
\section{Conclusions}

This paper introduces a novel problem, \textbf{\textit{vision to prediction of collective multi-agent dynamics}}. Multi-agent systems are defined as a collection of many ($>$10) interacting agents that form a collective dynamics. Our objective is to directly learn and predict collective dynamics (not states of each agent) of a multi-agent system from scenes, which will allow real-time perception and control of the system. We perform a comparative study between frame- and event-based methods, and empirically demonstrate the advantage of event representation in learning and predicting collective behavior of multi-agent systems. To the best of our knowledge, this is the first study to discuss multi-agent dynamic prediction from visual observation and demonstrating event-based method for it. We also present a new transformer-based deep learning architecture, evMAP, for better understanding of collective behavior of multi-agent system under dynamic changes by using adaptation of two memory latents.

\paragraph{\textbf{Acknowledgement}}
This work is supported by the Defense Advanced Research Projects Agency (DARPA) under Grant Numbers GR00019153. The views and conclusions contained in this document are those of the authors and should not be interpreted as representing the official policies, either expressed or implied, of the Department of Defence, DARPA, or the U.S. Government.     % 0.5p

\bibliographystyle{unsrt}  
\bibliography{references} 

\appendix

\clearpage
\section{Flocking Model: Reynolds Rule}

Craig W. Reynolds~\cite{reynolds1987flocks} has suggested three rules for the realistic modeling of flocking behaviors. 
Let's consider a system composed of \(N\) agents, where each agent \(i\) has a state defined by its position \(P_i(t)\) and velocity \(V_i(t)\) at time \(t\). The dynamics of the system, according to Reynolds rules, can be defined as follows:

1. \textbf{Alignment} \(A_i(t)\) involves agents adjusting their speed and direction to match those of their neighbors, creating synchronized and fluid group movement. Agent \(i\) aligns its velocity with the average velocity of its neighbors.
   \[A_i(t) = \frac{1}{|N_i|}\sum_{j \in N_i} V_j(t)\]

2. \textbf{Cohesion} \(C_i(t)\) drives agents to move towards the average position of their neighbors, promoting group unity and cohesive movement. Agent \(i\) moves towards the average position of its neighbors.
   \[C_i(t) = \left( \frac{1}{|N_i|}\sum_{j \in N_i} P_j(t) \right) - P_i(t)\]

3. \textbf{Separation} \(S_i(t)\) focuses on maintaining sufficient distance among agents to prevent collisions and overcrowding, ensuring individual safety and personal space. Agent \(i\) keeps a distance from its neighbors to avoid crowding.
   \[S_i(t) = -\sum_{j \in N_i} \frac{P_j(t) - P_i(t)}{\|P_j(t) - P_i(t)\|^2}\]

where \(N_i\) denotes the set of neighbors for agent \(i\).

These behaviors adjust \(P_i(t)\) and \(V_i(t)\) over time. The dynamics of \(P_i(t)\) and \(V_i(t)\) can be captured by the equations:
\[P_i(t+\Delta t) = P_i(t) + V_i(t) \cdot \Delta t\]
\[V_i(t+\Delta t) = V_i(t) + \left( A_i(t) + C_i(t) + S_i(t) \right) \cdot \Delta t\]
where \(A_i(t)\), \(C_i(t)\), and \(S_i(t)\) are the contributions to the velocity change from alignment, cohesion, and separation, respectively. Understanding the system dynamics can be seen as understanding \(P_i(t)\) and \(V_i(t)\).

\begin{figure}[hb]
  \centering
  % \vspace{-0.1in}
   \includegraphics[width=0.7\linewidth]{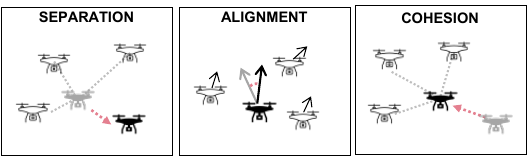}
   % \vspace{-0.1in}
   \caption{ \textbf{Three basic rules of flocking behavior.} Each agent avoids repulsion from other agents (\textit{separation}), steers towards the average heading of neighbors (\textit{alignment}), and steers towards the average position of neighbors (\textit{cohesion})~\cite{reynolds1987flocks}.}
   \label{fig:interactions_app}
   % \vspace{-0.1in}
\end{figure}
\clearpage
\section{Dataset Generation}

\subsection{Flocking System Simulation}
Based on Reynolds rules, we utilize flocking simulators to generate frames capturing various collective behaviors emerging from multi-agent interactions. One sequence contains 200 frames. 594 training sequences, 149 validation sequences, and 247 testing sequences are generated, each with various level of interactions among agents. These frames are used for training and evaluation of frame-based prediction models.

\paragraph{\textbf{NetLogo}}
% parameter, sample image
We utilize Flocking model provided by NetLogo~\cite{netlogoflocking} for 'Agent-wise Interaction Strength Detection' and 'Time Prediction of Collective Behavior Emergent' experiments. Several hyper-parameters are set as follows: max-cohere-turn 3, minimum-separation 0.5, vision 5, max-separate-turn 1.5, population 100. \textit{max-align-turn} varies in range of 1.5-5.5 and is used as the parameter to control the agent-wise interaction strength.

\paragraph{\textbf{AgentPy}}
% parameter, sample image
We utilize Flocking model provided by AgentPy~\cite{foramitti2021agentpy} for 'System Dynamic Change Detection' experiment. Several hyper-parameters are set as follows: inner-radius 3, outer-radius 10, cohesion-strength 0.005, seperation-strength 0.1, alignment-strength 0.3, border-strength 0.5, population 100. cohesion-strength, seperation-strength, alignment-strength are set to 0 for non-interaction cases.

\subsection{Vision to Event Conversion Framework}
The frames generated from flocking model are converted into event data using \textbf{v2e Framework}~\cite{hu2021v2e} for training and evaluation of event-based prediction models. This framework synthesizes realistic event data from intensity frames, accounting for non-idealities in event cameras, such as noise and motion blur. DVS 128 is considered with 1ms dvs exposure time and 33.3 input-slowmotion-factor. The generated event data contains about 2800 events/ms.

\subsection{Convergence Time}
We mathematically define the \textit{convergence-time} as the time point at which most agents are heading to a certain direction.
We define the \textit{convergence-time} as the time point at which the standard deviation of the min-max normalized sine and cosine values of the agents' heading directions drops below 0.5 as described below: 
\vspace{-0.1in}
\begin{multline}
T_c = \min \left\{ t \mid \text{std}(\|\text{min-max}(\sin \mathbf{\theta}_i(\tau))\|) < 0.5 \ \text{and} \right. \left. \text{std}(\|\text{min-max}(\cos \mathbf{\theta}_i(\tau))\|) < 0.5 \ \text{for all} \ \tau \geq t \right\}
\end{multline}
This definition is used to label convergence-time during the dataset generation. It offers a precise and quantifiable method for determining when the agents in the simulation have achieved a significant level of alignment in their movements, marking a key moment in the collective behavior of the group.
\clearpage
\section{Experiment Details}
% evmap detail
\subsection{Prediction Models}
\textbf{SlowFast\_8x8\_R50}~\cite{feichtenhofer2019slowfast}, \textbf{MoViT\_B\_16x4}~\cite{fan2021multiscale}, \textbf{MoViTv2\_S\_16x4}~\cite{li2022mvitv2} are trained based on the default settings provided in the official github repository of \cite{li2022mvitv2}. \textbf{AEGNN}~\cite{schaefer2022aegnn} is also trained based on the default settings provided in its official github repository. \textbf{Eventformer}~\cite{kamal2023associative} is trained with settings provided in the paper, with $32 \times 32$ memory dimension. For all these models, we replace their classification heads with regression heads, retraining them with L1 loss specifically for predicting interaction intensity and convergence-time.

\textbf{evMAP} is trained with L1 loss using Adam~\cite{kingma2014adam} with batch size 64 with initial learning rate of $1e^{-3}$ that decreases by a factor of 5 after every 25 epochs. $32 \times 32$ memory dimension is used for both short/long-term memories.

\section{Experiments}
\subsection{Agent-wise Interaction Detection}
\begin{figure}[hb]
  \centering
   % \vspace{-0.2in}
   \includegraphics[width=0.7\linewidth]{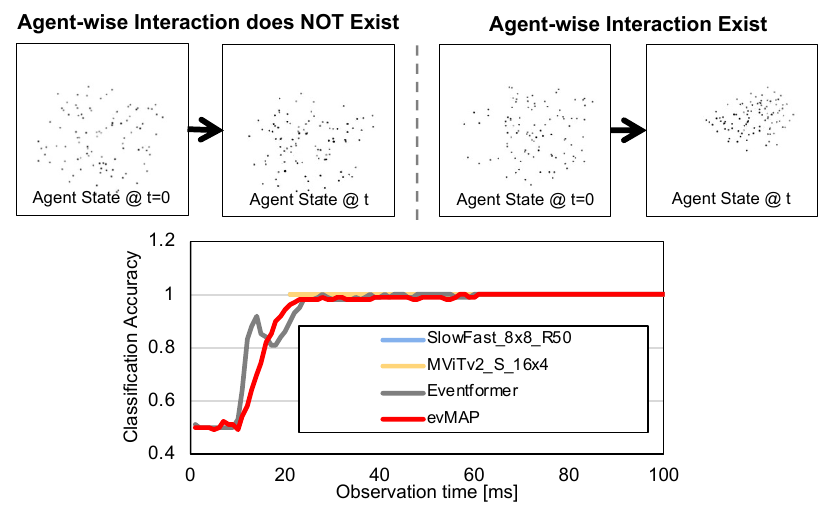}
    % \vspace{-0.2in}
   \caption{\textbf{Multi-agent behaviors} (top left) without interactions and (top right) with interactions. (bottom) \textbf{Average classification accuracy} from various prediction models. SlowFast's results are overlapped with MViTv2's.}
   % \vspace{-0.2in}
   \label{fig:swarm_cls}
\end{figure}

We first train prediction models to detect agent-wise interactions, using agent behaviors obtained from AgentPy~\cite{foramitti2021agentpy} for training and evaluation (Fig.~\ref{fig:swarm_cls}(left, middle)). All models use a classification head with 2 classes and are trained with cross-entropy loss. Simulation results indicate that all models achieve high accuracy (>90\%) within a short period, as shown in Fig.~\ref{fig:swarm_cls}(right). In particular, frame-based methods (SlowFast, MViTv2) reach an accuracy of 1 after the initial observation window. This is because agent-wise interactions result in a high density of agents, making them easier to identify from spatial information alone, thus diminishing the advantage of event-based representations with high temporal resolution.

\clearpage
\subsection{Agent-wise Interaction Strength Detection}

\begin{figure}[h!]
  \centering
   % \vspace{-0.2in}
   \includegraphics[width=0.7\linewidth]{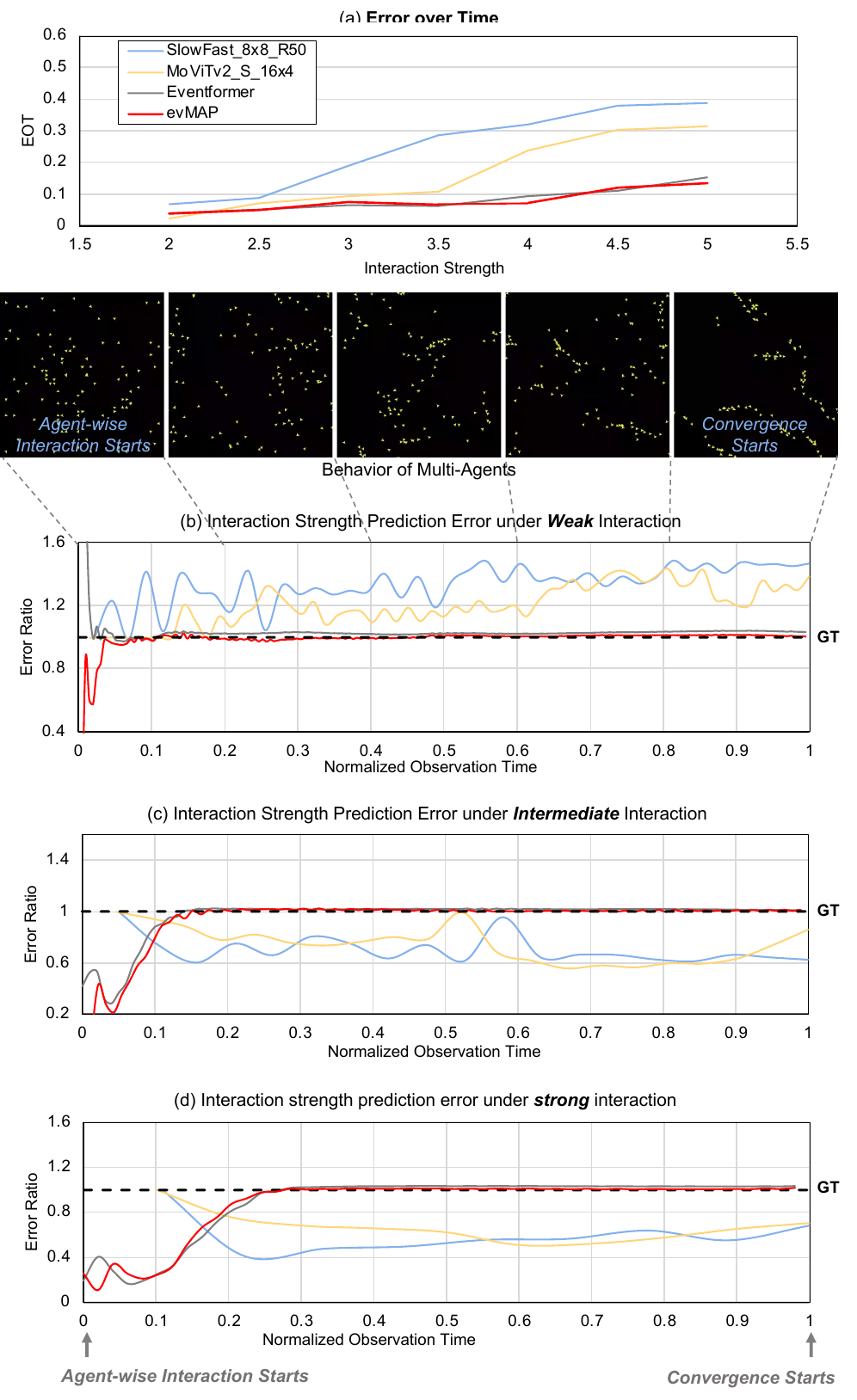}
    % \vspace{-0.2in}
   \caption{(a) \textbf{EOT of interaction strength prediction} across weak to strong interactions, (b-d) \textbf{Interaction strength prediction errors} from various frame- and event-based models (interaction strength=1.5, 3, 4.5). }
   % \vspace{-0.2in}
   \label{fig:error_turn_app}
\end{figure}

We present additional experiments on the detection of agent-wise interaction strengths across varying levels of interaction strength in Fig.\ref{fig:error_turn_app}. These results indicate that, with the exception of Eventformer\cite{kamal2023associative} and evMAP, prediction models struggle particularly with strong interactions.

\clearpage
\subsection{Time Prediction of Collective Behavior Emergent}

\begin{figure}[h!]
  \centering
   % \vspace{-0.2in}
   \includegraphics[width=0.7\linewidth]{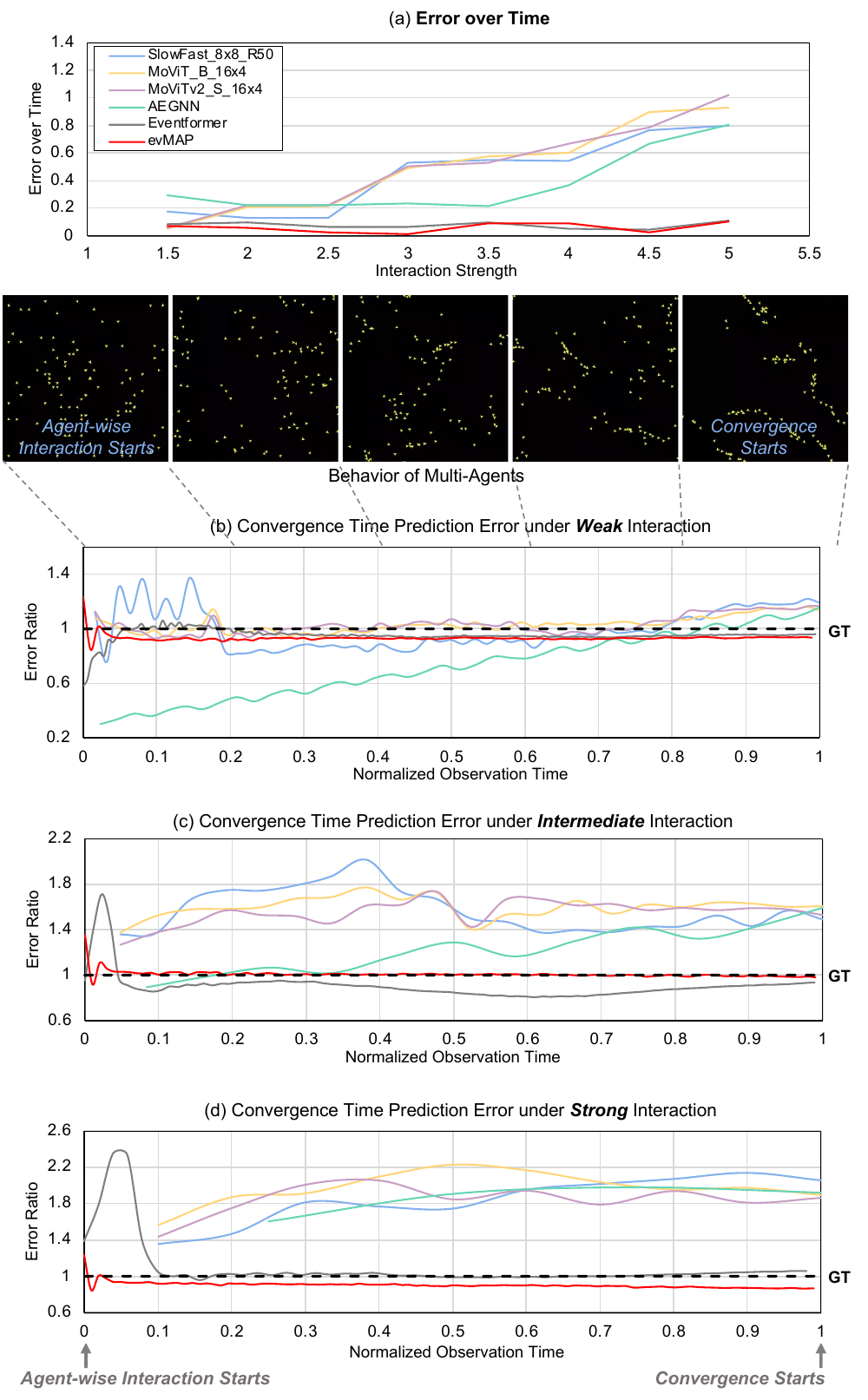}
    % \vspace{-0.2in}
   \caption{(a) \textbf{EOT for convergence-time prediction} across weak to strong interactions. (b) \textbf{Convergence-time prediction errors} from various frame- and event-based models (interaction strength=1.5, 3, 4.5).}
   % \vspace{-0.2in}
   \label{fig:error_time_app}
\end{figure}

We present additional experiments on the time prediction of collective behavior emergent across varying levels of interaction strength in Fig.\ref{fig:error_time_app}. These results indicate that, with the exception of Eventformer\cite{kamal2023associative} and evMAP, prediction models struggle particularly with strong interactions.

\clearpage
\section{Ablation Studies}
We present some ablation studies on different agent speeds and noise level of events.

\subsection{Agent Speed}
Event data is highly dependent on object speed; thus, evMAP predictions of convergence time under various agent speeds are explored in Table~\ref{tab:agent_speed}. Lower speeds result in fewer events, leading to increased EOT. However, these increased EOTs are still significantly lower than those from other frame-based methods (0.491-0.562).

\begin{table}[h]
  \centering
      \centering
      \caption{Different Agent speeds}
      \label{tab:agent_speed}
      \begin{tabular}{@{}ccc@{}}
        \toprule
        Norm. Speed & Avg. \# ev/ms & Avg. EOT$\downarrow$\\
        \midrule
        1 & 1506 & 0.055 \\
        0.5 & 1458 & 0.12 \\
        0.3 & 1289 & 0.15 \\
        \bottomrule
      \end{tabular}
\end{table}

\subsection{Real World Emulation with Noisy Events}
Real world data often includes events generated by objects or backgrounds unrelated to the agents-of-interest. We examine evMAP predictions of convergence time with noisy events in Table~\ref{tab:noise}. Noisy events lead to increased EOT due to the higher number of events, leading to increase in computations.

\begin{table}[h]
  \centering
      \centering
      \caption{Noisy Events}
      \label{tab:noise}
      \begin{tabular}{@{}ccc@{}}
        \toprule
        Noisy ev Ratio & Avg. \# ev/ms & Avg. EOT$\downarrow$\\
        \midrule
        0 (clean) & 1506 & 0.055 \\
        10\% & 1657 & 0.083 \\
        20\% & 1807 & 0.14\\
        \bottomrule
      \end{tabular}
\end{table}

\end{document}